\documentclass[manuscript]{aastex}
\usepackage{emulateapj5,natbib}

\slugcomment{accepted by the Astrophysical Journal}

\shortauthors{McCRADY ET AL.}
\shorttitle{SUPER--STAR CLUSTER M82-F}

\begin{document}

\title{ MASS SEGREGATION AND THE INITIAL MASS FUNCTION OF SUPER--STAR
CLUSTER M82-F\altaffilmark{1} }

\author{
Nate McCrady\altaffilmark{2,3}, James R. Graham\altaffilmark{3} and William D. Vacca\altaffilmark{4}
}
\altaffiltext{1}{
Based on observations made at the W.M. Keck Observatory, which is operated
as a scientific partnership among the California Institute of Technology,
the University of California and the National Aeronautics and Space
Administration.  The Observatory was made possible by the generous
financial support of the W.M. Keck Foundation.}

\altaffiltext{2}{
nate@astro.berkeley.edu}

\altaffiltext{3}{
Department of Astronomy, University of California, Berkeley, CA 94720-3411}

\altaffiltext{4}{
NASA Ames Research Center, Moffett Field, CA, 94035}

%%%%%%%%%%%%
% Abstract %
%%%%%%%%%%%%

\begin{abstract} 

We investigate the Initial Mass Function and mass segregation in
super--star cluster M82-F with high resolution Keck/NIRSPEC echelle
spectroscopy.  Cross-correlation with template supergiant spectra
provides the velocity dispersion of the cluster, enabling measurement
of the kinematic (virial) mass of the cluster when combined with sizes
from NICMOS and ACS images.  We find a mass of $6.6 \pm 0.9 \times
10^5 M_{\odot}$ based on near-IR light and $7.0 \pm 1.2 \times 10^5
M_{\odot}$ based on optical light.  Using PSF-fitting photometry, we
derive the cluster's light-to-mass ($L/M$) ratio in both near-IR and
optical light, and compare to population synthesis models.  The ratios
are inconsistent with a normal stellar initial mass function for the
adopted age of 40--60 Myr, suggesting a deficiency of low-mass stars
within the volume sampled.  King model light profile fits to new
HST/ACS images of M82-F, in combination with fits to archival near-IR
images, indicate mass segregation in the cluster.  As a result, the
virial mass represents a lower limit on the mass of the cluster.
\end{abstract}


%%%%%%%%
% Body %
%%%%%%%%

\section{Introduction}
\label{intro}

Star formation in starburst galaxies can be resolved into young,
dense, massive ``super--star clusters'' (SSCs) that represent a
substantial fraction of new stars formed in a burst event
\citep{meurer95, zepf99}.  SSCs identified from the ground two decades
ago \citep[e.g.,][]{arp85,melnick85} have been spatially resolved in
the nearest cases by the {\it Hubble Space Telescope} (HST) during the
last decade using WFPC in visible light
\citep[e.g.,][]{o'connell94,whitmore95}.

Optical studies of SSCs in dusty star forming regions are hampered by
high extinction.  The development over the past decade of new high
spatial resolution near-infrared imaging and high spectral resolution
spectroscopic instruments, including NICMOS aboard HST and the NIRSPEC
spectrometer at the W. M. Keck Observatory, has led to a wealth of new
data on SSCs.  In the optical, the Advanced Camera for Surveys (ACS)
imager on HST has substantially improved resolution over the standards
set by WFPC2.  These new instruments have enabled direct measurement
of structural parameters and determination of virial masses for even
heavily reddened, optically-invisible clusters.

In addition to being a significant mode of star formation in starburst
galaxies, SSCs are important for study of the formation of high mass
stars, galaxy stellar populations and evolution, and physical
conditions analogous to high-redshift star formation.  Numerical
simulations by \citet{zwart04} further suggest that runaway mergers of
high mass stars deep in the potential well of dense SSCs lead to the
creation of intermediate-mass black holes, a possible missing link
between stellar mass black holes and the supermassive black holes
found in the centers of galaxies.  Especially important is the stellar
initial mass function (IMF), for which SSCs are ideal laboratories:
coeval populations with enough stars to extensively sample the IMF.
Critical to understanding the IMF is detection of low-mass stars, the
light of which is swamped by high luminosity supergiants.
Measurement of the kinematic mass of the clusters represents the only
means of detecting and quantifying the contribution of low-mass stars.

At a distance of $3.6 \pm 0.3$ Mpc \citep{freedman94}, M82 is the
nearest massive starburst galaxy.  \citet{lipscy04} find that at least
20\% of the star formation in M82 is occuring in SSCs.  The galaxy's
high inclination of 81$^\circ$ \citep{achtermann95} and prevalent dust
lead to large, patchy extinction; near-IR observations are required to
overcome this obstacle in characterizing the SSC population.  The
nuclear starburst in M82 contains over 20 SSCs that are prominent in
the near-infrared, with typical ages of $\sim 10^7$ years
\citep{natascha98}.  These young, massive clusters offer an excellent
opportunity to search for variations in the IMF.

Early studies \citep{rieke80, mcleod93} of the IMF in the M82
starburst used ground-based observations and were necessarily global
in scale; these studies indicated an abnormal IMF for the starburst.
An influential work by \citet{rieke93} used population synthesis
models to constrain the IMF.  They concluded that the large $K$-band
luminosity of the M82 starburst relative to the dynamical mass
required the IMF to be significantly deficient in low-mass stars ($M <
3M_{\odot}$) relative to the solar neighborhood.

Local examination of star formation regions on the scale of individual
SSCs provides an important test for the assertion of an abnormal,
``top-heavy'' IMF. \citet{satyapal97} used $1''$-resolution near-IR
images to identify pointlike sources and found that at this scale
starburst models could match observations without invoking a high-mass
biased IMF.  \citet{mccrady03} used high-resolution ($0.''2$)
HST/NICMOS images and near-IR Keck/NIRSPEC spectroscopy to measure the
kinematic (virial) masses of two individual near-IR-bright super--star
clusters in the nuclear starburst.  Comparison of derived
light-to-mass ratios and population synthesis models indicate that
cluster MGG-9 has a normal IMF, while MGG-11 may be deficient at the
1-$\sigma$ level in low-mass stars ($M<1$ M$_{\odot}$).  This work
indicates that the IMF may vary on the scale of individual clusters,
and bears on the ongoing debate over the role of physical conditions
in a star formation environment in determining the mass distribution
function of newly-formed stars.

The brightest optically-visible cluster within 500 pc of the center of
the galaxy is M82-F.  \citet[][hereafter SG01]{smith01} studied the
cluster in detail using optical spectra and HST/WFPC2 images to
estimate the cluster's virial mass.  Based on population synthesis
models, they determined that M82-F is too luminous for its assumed
age, and must therefore have a top-heavy IMF.  With most of its mass
in stars of $M > 2 $M$_{\odot}$ and a deficiency of long-lived,
low-mass stars, the authors suggested that the cluster is ``doomed''
and will begin to dissolve due to stellar evolution over the next 1--2
Gyr.  We examine some of the limitations of their data in Section
\ref{acsdata}.  SG01 noted the possibility that their data could
instead reflect mass segregation, which would lead to underestimation
of the cluster's mass.  Using NICMOS images, \citet{mccrady03}
measured a smaller radius for M82-F in near-IR images, suggesting that
observed light-to-mass ratios may be systematically overestimated due
to mass segregation.

Mass segregation cannot be observed directly via star counts for
clusters in which individual stars are unresolved.  However,
\citet{sternberg98} notes that a mass-segregated cluster should appear
smaller in the UV and IR (light dominated by red supergiant stars)
than in the optical (where the light is dominated by intermediate-mass
main sequence stars).

In this paper, we characterize the IMF of the super star cluster M82-F
and investigate indications of mass segregation.  We measure the
cluster stellar velocity dispersion using new high spatial resolution
Keck/NIRSPEC echelle spectra and present analysis of new,
high-resolution, multi-waveband optical images of the cluster from the
High Resolution Camera on the HST Advanced Camera for Surveys (ACS).
As discussed in Section \ref{analysis}, these new images represent a
significant improvement in the characterization of the physical
structure of M82-F.  We fit two-dimensional light profiles to the
cluster in the ACS and archival HST/NICMOS images, and derive the
kinematic mass and light-to-mass ratio.  Section \ref{obs} describes
the observations used in this work, Section \ref{analysis} presents
analysis of the data and determination of key cluster parameters, and
Section \ref{discussion} discusses mass segregation and the form of
the IMF in M82-F.

\vspace{0.5in}

\section{Observations}
\label{obs}

We observed super star cluster M82-F (MGG-F) with the 10-m Keck II
telescope on 2003 January 19, using the facility near-infrared echelle
spectrometer NIRSPEC.  We obtained high-resolution ($R \sim 22,000$),
cross-dispersed spectra in the wavelength range 1.51--1.75 $\mu$m
using the NIRSPEC-5 order-sorting filter.  The data fall in seven
echelle orders, ranging from $m=44$ through 50.  The cluster was
nodded along the $0.''432$ $\times$ $24''$ slit, with successive nods
separated by $\sim 10$ arcsec (seeing was $\sim 0.''8$).  Seven
separate integrations, each 600 seconds in duration, provided a total
time of 4200 seconds on the cluster.  The spectra were dark
subtracted, flat-fielded and corrected for cosmic rays and bad pixels.
The curved echelle orders were then rectified onto an orthogonal
slit-position/wavelength grid based on a wavelength solution from sky
(OH) emission lines.  Each pixel in the grid has a width of
$\delta\lambda = 0.019$ nm.  We sky-subtracted by fitting third-order
polynomials to the two dimensional spectra column-by-column.

The spectra were extracted using Gaussian weighting functions matched
to the wavelength-integrated cluster profile.  Atmospheric calibration
was performed using the spectrum of HD 173087, a B5V star.  The
calibration star was observed at an airmass of 1.94, consistent with
the airmass range (1.76--2.07) of the M82-F observations.  To account
for photospheric absorption features (particularly Brackett series and
helium lines) and continuum slope, the calibration star spectrum was
divided by a spline function fit.  The resulting atmospheric
absorption spectrum was divided into the cluster spectra to recover
absorbed flux.  The individual extracted spectra have an average
signal-to-noise ratio (S/N) of $\sim 25$ per pixel and S/N $\sim 66$
per pixel when all integrations are combined.  For comparison, the
optical spectra used in SG01 have S/N per pixel of 15--25.

For photometric measurements and structural parameters, we obtained
new high-resolution optical images of M82-F.  The images were taken
2002 June 7 with the ACS High Resolution Camera (HRC) in the F435W,
F555W and F814W filters (corresponding approximately to the Johnson
$B$, $V$ and $I$ filters).  Exposure times for the images are 1320
seconds, 400 seconds, and 120 seconds, respectively.  Additional
observations in the F250W filter were taken on 2003 Jan 20, with
exposure time of 10,160 seconds.  ACS/HRC has a plate scale of 25 mas
per pixel and is thus critically sampled at 477 nm.  We relied on
pipeline reduction (CALACS) for bias correction, dark subtraction and
flatfielding.  The resulting set of ``FLT'' images were processed with
the Multidrizzle package \citep{koekemoer02} to reject cosmic rays,
mask bad pixels and create a ``drizzled'' image for each filter,
corrected for geometric distortions.  

In the near-IR, we used archival HST/NICMOS images in the F160W and
F222M filters.  For details on these data and their reduction, see
\citet{mccrady03}.

\section{Analysis}
\label{analysis}

\subsection{Stellar Velocity Dispersion}

\begin{figure}
\centering \epsscale{0.95}
\plotone{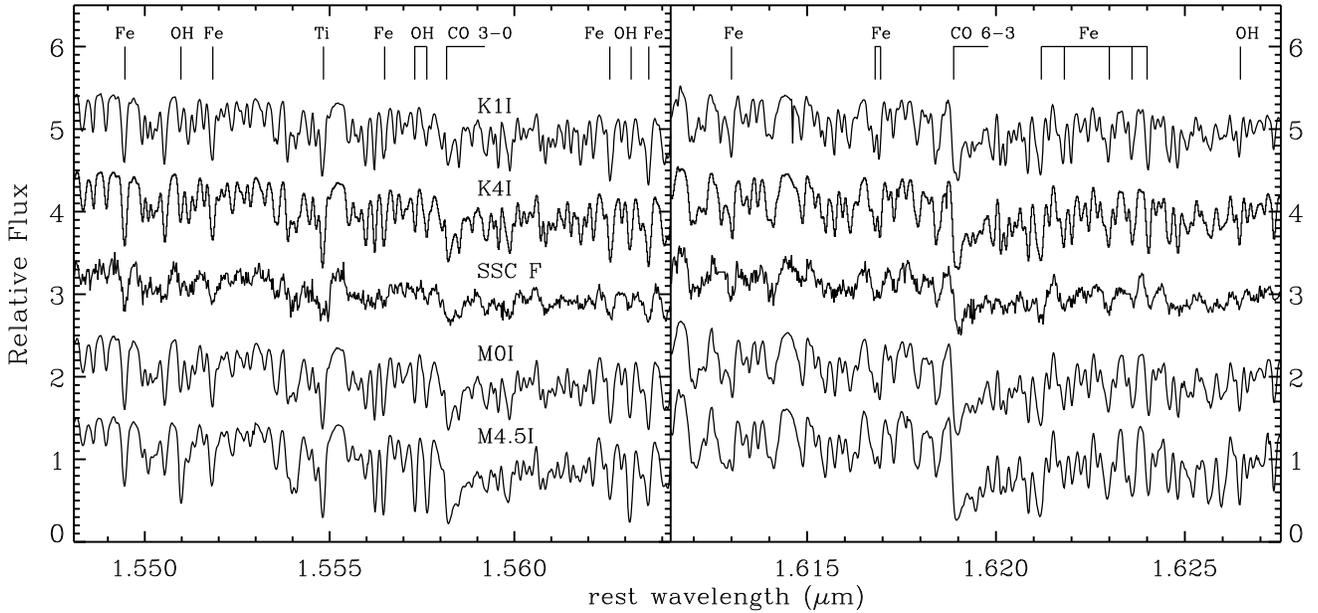}
\caption{Near-IR rest-frame spectra of echelle orders 49 and 47 for M82-F
and a range of template supergiants.  The spectra are identically
normalized, but offset vertically for clarity (zero points are $-$0.7,
0.3, 1.3, 2.3 and 3.3, from bottom to top).  The cluster spectrum
closely resembles the supergiant spectra, with the features ``washed
out'' by the stellar velocity dispersion (e.g., the CO bandheads at
$1.5582 \mu$m and $1.6189 \mu$m).}
\label{spect}
\end{figure}

\begin{figure}
\centering
\epsscale{0.4}
\plotone{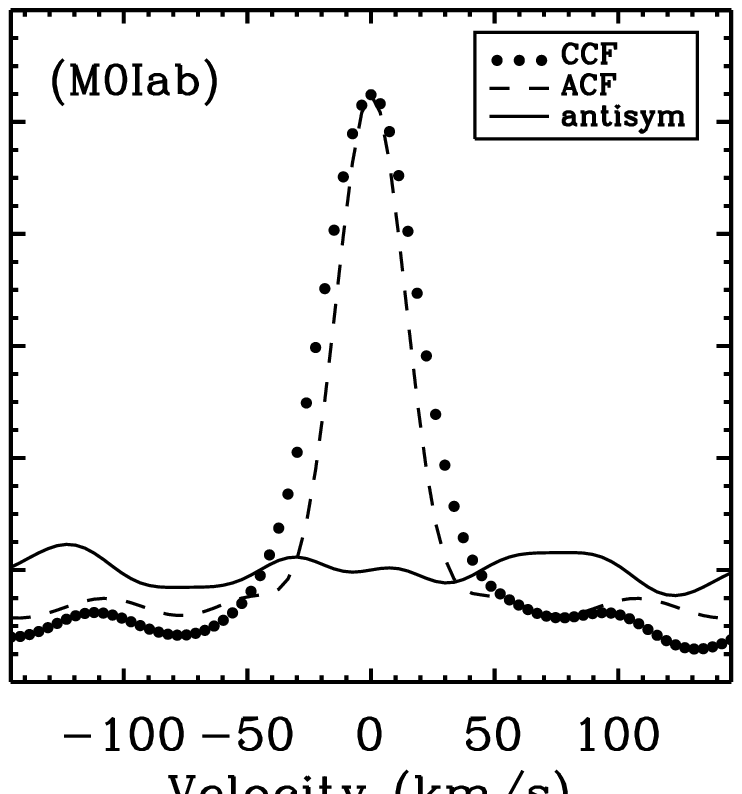}
\caption{Cross-correlation function (CCF) for the cluster and the
template M0Iab star, averaged over all echelle orders.  The peak of
the template star auto-correlation function has been scaled to the
peak of the CCF to emphasize the difference in width (due to the
convolution with the velocity distribution function of the cluster).
The solid line is the antisymmetric part of the CCF, which is seen to
be highly symmetric. }
\label{ccfplot}
\end{figure}

To derive the kinematic mass of a cluster using the virial theorem, we
need a measurement of the internal velocity dispersion in the cluster
and a characteristic radius.  Figure \ref{spect} shows a portion of
the NIRSPEC echelle spectrum of M82-F, compared to a series of
template supergiant spectra.  Features found in the spectra of
supergiant stars are readily identified in the cluster spectrum,
although they appear ``washed out'' due to the stellar velocity
dispersion.  Especially prominent are the rovibrational CO bandheads
and numerous Fe and OH lines.  We have assembled an atlas of
high-resolution NIRSPEC spectra of supergiant stars for use in
cross-correlation analysis \citep{mccrady03}, from which we determine
the dominant spectral type and line-of-sight velocity dispersion.
Based upon the peak of the cross-correlation function (Figure
\ref{ccfplot}), the $H$-band spectrum of M82-F most closely matches
spectral types in the range of K4I--M0I.  The line-of-sight velocity
dispersion based on cross-correlation with templates in this range is
$\sigma_r = 13.5 \pm 0.2$ km s$^{-1}$.  The value of $\sigma_r$
decreases as a function of the similarity to the template spectrum as
measured by the peak value of the cross-correlation function.  It is
therefore possible that the stated value reflects some template
mismatch bias due to our limited template spectra atlas, but this
effect is small ($<$ 0.5 km s$^{-1}$).

\subsection{Cluster Size and Mass}

\begin{figure}
\centering \epsscale{0.5} \plotone{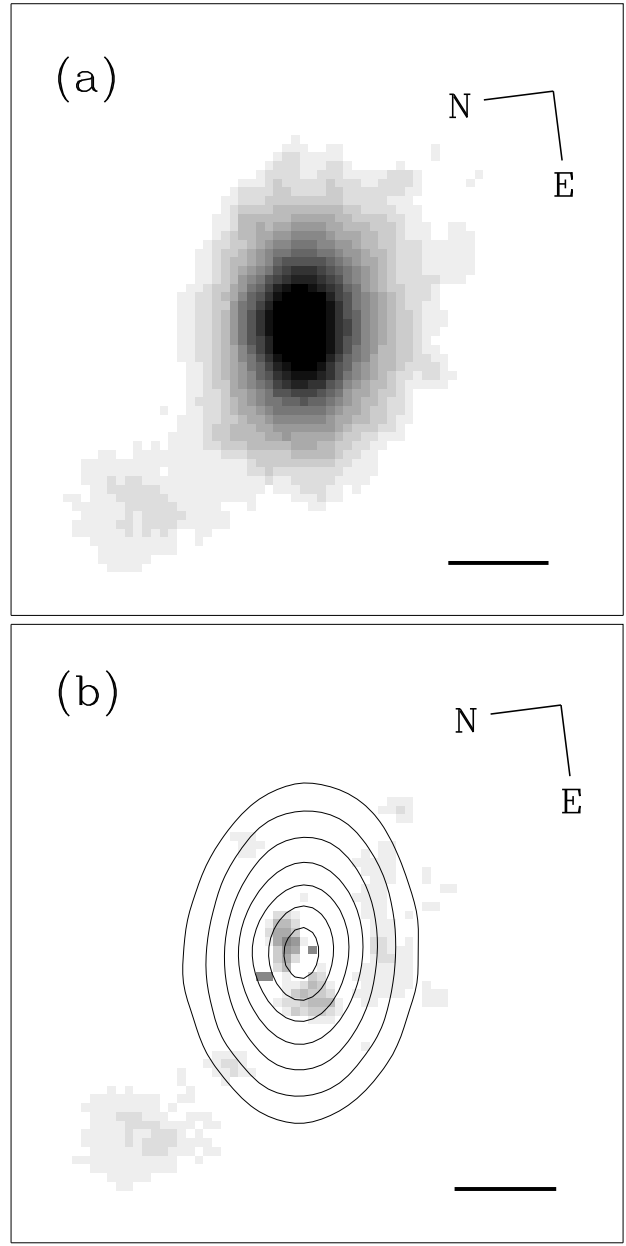}
\caption{Comparison of data and residuals from the fitted
two-dimensional King function.  Plot (a) shows the ACS/F814W image of
M82-F, logarithmically scaled to accentuate faint features.  Plot (b)
shows the data image minus the fitted King function, with the same
scaling as plot (a).  Contours show the fit to the cluster, at
1,2,4,8,16,32 and 64 percent of the peak value.  Scale bars in the two
images represent 5 pc ($0.''29$) at the adopted distance of M82.  }
\label{fitplot}
\end{figure}

The simplest approach to determine the cluster mass is to measure the
half-light radius, assume that light traces mass, the cluster is
spherical, and the velocity dispersion is isotropic, then apply the
virial theorem.  M82-F presents a difficulty for this method: HST
images (Figure \ref{fitplot}) clearly show that the cluster is
elliptical in projection and cannot, therefore, be spherical.  To
measure the radius, we fit the cluster using an elliptical version of
the empirical \citet{king62} model:

\begin{equation}
f(x,y) = k_1 \left\{ \left( 1+\frac{x^2}{a^2}+\frac{y^2}{b^2}\right)^{-1/2}
 - (1+k_2^2)^{-1/2} \right\}^2
\end{equation}

\noindent where $a$ and $b$ are the characteristic lengths of the
minor and major axes, respectively, $k_1$ is a scaling constant and
$k_2$ acts as a ``tidal radius,'' truncating the profile beyond a
particular scale length.  The fit therefore has four free parameters
($k_1$, $k_2$, $a$ and $b$) to describe the light profile and three
more to describe the centroid location and position angle.  The King
model is convolved with a model PSF from Tiny Tim \citep{krist95} and
compared to the image.  Fit parameters are determined by iterative
search over parameter space, using a Levenburg-Marquardt least-squares
fit.  Figure \ref{fitplot} shows the fit and residuals for the ACS/HRC
F814W image of M82-F.

The half-light radius in projection along the major axis, $r_{hp}$, is
the semimajor axis of the ellipse that encloses half the flux in the
fitted King model.  We determine $r_{hp}$ numerically by summing the
flux in a series of ellipses with the axial ratio defined by $a$ and
$b$.  Monte Carlo simulations of clusters indicate that $a$, $b$ and
$k_2$ are significantly covariant; however, the fitted projected
half-light radius along a given axis is accurate to about 2 percent.
The half-mass radius may be determined by assuming that light traces
mass and deprojecting by dividing $r_{hp}$ by 0.766 \citep{spitzer87}.

\citet{mccrady03} fit a spherical King model to M82-F in the NICMOS
F160W image and found a projected half-light radius of $89 \pm 11$
mas.  Our elliptical King function fit to the same image found
$r_{hp}=113 \pm 2$ along the major axis and $r_{hp}=62 \pm 1$ along
the minor axis.  At the adopted distance of M82, the projected
half-light radii along the major and minor axes are thus $1.97 \pm
0.16$ pc and $1.1 \pm 0.1$ pc, respectively.

\begin{figure}
\centering
\epsscale{0.4}
\plotone{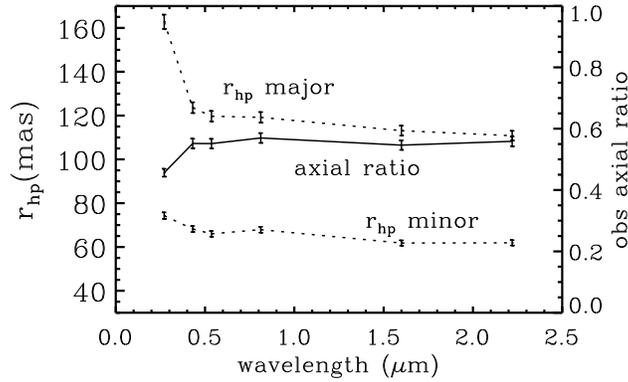}
\caption{Measured half-light radius and observed axial ratio as a
function of wavelength.  Lines are presented as a guide to the eye.
The general decrease in size with wavelength is consistent with mass
segregation.  Note that the axial ratio remains constant within the
error for all wavelengths other than 250 nm.}
\label{halfradii}
\end{figure}

To account for the ellipticity of the cluster, we assume it is an
oblate spheroid and compute the gravitational potential assuming that
the cluster is homoeoidal \citep[][Section 2.3]{binney87}.  The
numerical constant in the virial mass formula may be separated into
factors dependent upon the central concentration and the ellipticity.
The compact core plus extended envelope structure of star clusters is
similar to an $n=5$ polytrope \citep[][p. 13]{spitzer87}.  For an
$n=5$ polytropic oblate ellipsoid with an isotropic velocity
dispersion, the virial mass is:

\begin{equation}
M = 10.31 \, (e/\sin^{-1}{e}) \, r_{hp} \, \sigma_r^2 /G
\end{equation}

\noindent where the eccentricity is $e = \sqrt{1-(Z/R)^2}$, and $R$
and $Z$ are the equatorial and polar radii of the oblate spheroid,
respectively.  Figure \ref{halfradii} shows the fitted cluster
half-light radii in projection along the major and minor axes for the
NICMOS and ACS images.  The observed axial ratio, $a/b$, represents a
lower bound on the ratio $Z/R$; an oblate cluster will appear rounder
than its intrinsic shape unless viewed directly along the equatorial
plane.  Averaging over all possible inclinations, we find that the
observed axial ratio of $0.55$ corresponds to a most-likely intrinsic
axial ratio of $0.35$ with $e=0.77 \pm 0.07$.  We represent the
uncertainty on the viewing inclination as the difference between the
angle-averaged value and the lower bound.

We believe the ellipticity of the cluster is intrinsic, rather than
the result of differential reddening.  Extinction is very low at 2.2
$\mu$m, and therefore the observed shape in the F222M images is very
likely intrinsic.  Moreover, the axial ratio is constant within the
uncertainties across all wavebands from 0.4 to 2.2 $\mu$m (Figure
\ref{halfradii}).  If the ellipticity were due to the distribution of
dust around the cluster, the axial ratio should change as a function
of wavelength as the extinction is expected to vary significantly
between $B$ and $K$.  The axial ratio does appear to increase sharply
at the shortest HST/ACS waveband (F250W).  We interpret this as a
result of scattering of cluster light by dust outside the cluster.
The ACS images show strong spatial variations in extinction in the
immediate vicinity of M82-F.  Scattering and absorption in the
ultraviolet compromise the quality of the fit to the light profile at
very short wavelengths.

Combining the NIRSPEC velocity dispersion with the NICMOS radius and
estimated eccentricity, we calculate an $H$-band virial mass of $M_H =
6.6 \pm 0.9 \times 10^5$ M$_{\odot}$ for M82-F.  The subscript on the
mass is used to emphasize that the virial mass is computed from a
velocity dispersion and size measured in the $H$ band.  In this manner
we minimize systematic errors by measuring both variables from the
light of the same stars and may seek variations between different
wavebands.

\subsection{Optical Results \& Photometry}

SG01 found a velocity dispersion of $13.4 \pm 0.7$ km s$^{-1}$ for
M82-F based on cross-correlation analysis using cluster spectra in the
601--759 nm range (overlapping the ACS F814W bandpass).  They
concluded that the cluster spectrum most closely matches their
template K stars, consistent with the Starburst99 prediction that K II
stars dominate the spectrum at the inferred cluster age (discussed in
Section \ref{age}).  We adopt these results to measure the cluster
mass with optical data.

In the ACS F814W image, we measure a half-light radius of $119 \pm 2$
mas.  Using this value and the SG01 velocity dispersion, we find an
$I$-band virial mass of $M_I = 7.0 \pm 1.2 \times 10^5$ M$_{\odot}$
for M82-F.  This is significantly lower than the optically-derived
SG01 mass estimate of $1.2 \pm 0.1 \times 10^6$ M$_{\odot}$ due to
more accurate measure of the cluster's half-light radius (see Section
\ref{acsdata}).

Photometry was derived by integrating over the fitted King models and
applying the appropriate conversion factors for each instrument and
filter.  Details of the PSF fitting photometry and spectral energy
distribution are presented in a companion paper \citep{vacca04}.  The
observed cluster in-band luminosities as defined in \citet{mccrady03},
not corrected for extinction, are $L_{F814} = 3.8 \pm 0.6 \times 10^5$
L$_{\odot}$ and $L_{F160} = 4.0 \pm 0.7 \times 10^5$~L$_{\odot}$.

\section{Discussion}
\label{discussion}

By comparing the derived light-to-mass ($L/M$) ratio of M82-F to
population synthesis models, we may characterize the IMF of the
cluster.  Critical to this analysis are the cluster age and
line-of-sight extinction.  

\subsection{Age}
\label{age}

The spectrum of M82-F immediately places upper and lower bounds on the
cluster's age.  There is no evidence of nebular emission in the $H$-
or $K$-band, which indicates the absence of O stars and a minimum
cluster age of 6--7 Myr.  \citet{natascha98} finds none of the
features expected of AGB stars for the nuclear region of M82, setting
an upper limit of $\sim 10^8$ years.  For more precise limits on the
age, we turn to population synthesis modelling.  As noted in Section
\ref{analysis}, the F160W light of M82-F most closely matches template
spectral types in the range of K4I--M0I.  \citet{gilbert02th} used
Starburst99 population synthesis models to determine the flux-weighted
average spectral type as a function of age for coeval stellar
populations.  In the $H$-band, the cluster light is dominated by
K4--M0 stars for a brief period around 15 Myr and during the ages of
$\sim$ 40--60 Myr.  SG01 used Starburst99 models to fit the H$\delta$
and He I absorption profiles in optical spectra.  They found that the
best fits to the wings of the Balmer line and depth of the helium line
suggest an age of $60 \pm 20$ Myr.  This is consistent with the
dominant spectral type of the $H$-band light, and we therefore adopt
an age range of 40--60 Myr for M82-F.

\subsection{Extinction}
\label{extinction}

SG01 used $BVI$ photometry to determine line-of-sight extinction of
$A_V = 2.8$.  Applying the \citet{cardelli89} extinction law ($R_V =
3.1$) to the this value gives $A_{F160W} = 0.53$ and $A_{F814W} =
1.63$.

As an independent test, we calculate synthetic colors for the near-IR
dominant evolved K4--M0 stars based on the \citet{pickles98} stellar
spectral library.  The synthetic [F160W] $-$ [F222M] color for a K4I
star is 0.35, versus 0.52 for M0I.  From our photometry, we find
[F160W] $-$ [F222M] = $0.36 \pm 0.04$ for M82-F, at the blue end of
the expected color range.  This implies that $H$-band extinction to
M82-F is quite small, essentially negligible.

The optical (600-800 nm) light of the cluster is expected to be
dominated by KII stars (SG01).  Synthetic [F555W] $-$ [F814W] colors
for these stars in the Pickles library range from 1.19 to 1.45, versus
the measured $1.71 \pm 0.04$ for M82-F.  For a standard ($R_V = 3.1$)
interstellar extinction curve, this color excess gives $A_{F814W} =
0.34$ to 0.66 mag.

\begin{figure}
\centering
\epsscale{0.4}
\plotone{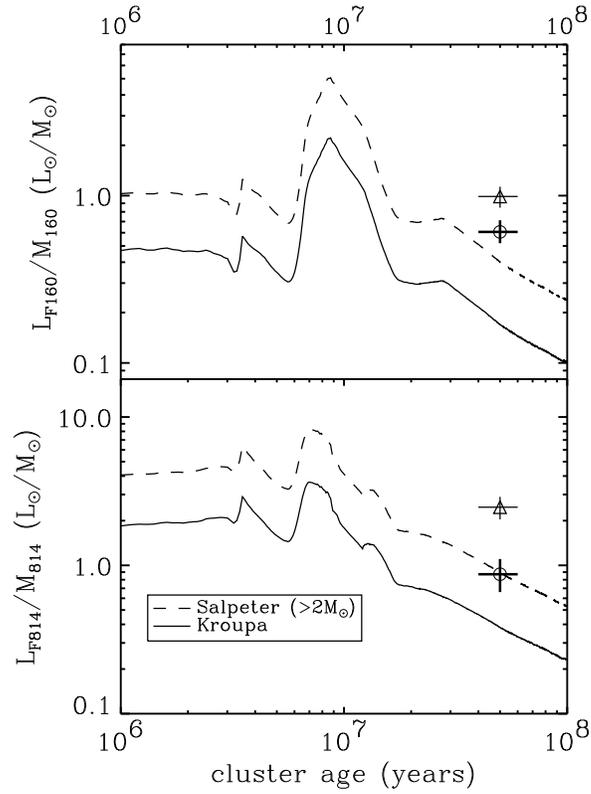}
\caption{Comparison of the derived light-to-mass ratios with
predictions from Starburst99 models for two IMFs.  The L/M ratio for
the cluster is clearly inconsistent with a normal Kroupa IMF.
$H$-band is shown in the top plot and $I$-band is shown in the bottom
plot.  Circles mark L/M based on extinction estimates derived from
synthetic photometry (Section \ref{extinction}) while triangles mark
values based on the SG01 extinction estimate.}
\label{sb99plot}
\end{figure}

To calculate the light-to-mass ratio, we need to deredden the cluster.
We adopt extinction of $A_{F160W} = 0.0 ^{+0.1}_{-0.0}$ and $A_{F814W}
= 0.5 \pm 0.2$ based on the synthetic photometry results.  For sake of
comparison, Figure \ref{sb99plot} reflects both these estimates and
the extinctions implied by the SG01 estimate of $A_V$.

The Galactic dust map of \citet{schlegel98} indicates Galactic
extinction along the line of sight to M82-F of $A_V = 0.48$,
$A_{F814W} = 0.28$ and $A_{F160W} = 0.09$.  This provides a
cross-reference for the estimated extinction.

\subsection{The IMF}

We determine the luminosity and virial mass of the cluster independent
of any assumptions about the IMF.  This is in contrast to photometric
mass determinations, which are based on cluster colors and ages and
must therefore assume an IMF.  Our method enables us to constrain the
cluster IMF by comparing observed light-to-mass ratios in various
wavebands to population synthesis models \citep{sternberg98,mccrady03}.

By applying the adopted extinction corrections for M82-F, we find
de-reddened light-to-mass ratios of $L/M = 0.6 \pm 0.1$
(L$_\odot$/M$_\odot$) at 1.6$\mu$m and $L/M = 0.9 \pm 0.2$ at
0.8$\mu$m.  Figure \ref{sb99plot} compares the measured $L/M$ ratio to
population synthesis model predictions for two fiducial IMF forms.  We
used Starburst99 version 4.0, with an instantaneous burst, solar
metallicity \citep{mcleod93} and the Hillier \& Pauldrach atmosphere
models.  The derived $L/M$ ratio of M82-F is too high for the standard
\citet{kroupa01} IMF over the full range of stellar masses (0.1--100
M$_{\odot}$).  Rather it appears that the cluster's IMF is deficient
in low-mass stars.  For example, the $L/M$ ratio is roughly consistent
with a \citet{salpeter55} IMF truncated at a lower mass of about 2
M$_{\odot}$.  If the cluster age were approximately 15 Myr, the
derived $L/M$ ratios would be consistent with the Kroupa IMF; an
independent determination of the cluster's age could rule out this
possibility.

In our analysis we assume a particular form for the IMF (i.e., Kroupa
or Salpeter power laws) and modify the lower-mass cutoff to generate
the observed $L/M$ ratio at the adopted age.  This method allows us to
test the top-heavy IMF hypothesis --- if the observed $L/M$ ratio is
fit with an IMF extending down to 0.1 M$_{\odot}$, as in cluster MGG-9
in \citet{mccrady03}, there is no need to invoke an abnormal IMF.  In
the present case, a normal (Kroupa) IMF cannot explain the relatively
large $L/M$ ratio of the cluster.  An elevated lower-mass cutoff is
not required, however.  The observed $L/M$ ratios could be caused at
the cluster's age by flattening the slope of the IMF, which would
change the relative proportions of stellar masses.  Although our data
do not distinguish between changes to the lower mass cutoff or IMF
slope, it is evident that the IMF is different than the Kroupa form,
and different from other nearby SSCs in the M82 nuclear starburst.

A similar study has been conducted by \citet{mengel02} in the Antennae
(NGC 4038/4039).  They determined virial masses and light-to-mass
ratios for a sample of five bright SSCs, and compared the results to
various IMF forms using population synthesis models.  The clusters
studied by Mengel et al. appear to exhibit a range of IMFs, with some
evidence of dependence on location within the merger environment.  In
contrast, \citet{larsen04} measured virial masses for five clusters in
nearby galaxies and found light-to-mass ratios consistent with a
Kroupa or Salpeter IMF.  They found no evidence for a deficiency of
low-mass stars in these clusters.

\subsection{Optical Images}
\label{acsdata}

The ACS observations, designed specifically to study the properties of
M82-F, represent a significant advance in quality over all earlier
optical images.  \citet{o'connell95} imaged the cluster with the
pre-repair mission Planetary Camera in 1992.  The cluster fell at the
edge of or between chips on the detector in their $V$- and $I$-band
equivalent images, and they deemed their $V$-band deconvolution ``not
highly reliable.''  SG01 used archival WFPC2 images in the F439W,
F555W and F814W filters \citep{degrijs01}.  These images were designed
for a separate study of M82, and were not ideal for M82-F.  The
cluster fell on the WF4 CCD, and were significantly undersampled due
to the 100-mas pixels.  The two longer-wavelength images were both
saturated in the cluster core.  The ACS/HRC images used exposure times
specifically designed for study of M82-F, and with 25 mas pixels, are
critically sampled above 477 nm.  SG01 did not deconvolve the PSF from
the WFPC2 F439W image, instead estimating the broadening
heuristically.  By contrast, we treated the PSF more rigorously using
a model PSF as described in Section \ref{obs}.  Both the
\citet{o'connell95} and SG01 studies found a projected half-light
radius of $160 \pm 20$ mas for M82-F, using images at 555 nm and 439
nm, respectively.  Our fits to the major axis of the cluster in
ACS/HRC images yield significantly smaller projected half-light radii
of $120 \pm 2$ mas and $123 \pm 2$ mas at 555 nm and 435 nm,
respectively.  These values are more precise and more accurate than
the radii determined from the lower-resolution optical data in earlier
works.

\subsection{Mass Segregation}

The axial ratio of M82-F is $\sim 0.55$ in all but the shortest
wavelength image (Figure \ref{halfradii}).  Fits to the images show a
distinct trend of decreasing cluster size with increasing wavelength.
Light at shorter wavelengths is increasingly dominated by hotter stars
still on the main sequence.  In a coeval population, these stars will
be intermediate mass stars.  Longer wavelength light is dominated by
cooler evolved stars, stars originally more massive than those at the
cluster's main sequence turnoff point.  The negative correlation
between cluster size and observed wavelength suggests that the massive
red evolved stars that dominate the near-IR light are more centrally
concentrated than the intermediate-mass main sequence stars which
dominate the optical light \citep{sternberg98}.

A key assumption of our method is that light traces mass within the
cluster.  Mass segregation renders the interpretation of the
light-to-mass ratio problematic.  The near-IR measurements presented
here trace the light of supergiant stars --- the most massive stars
currently present in the cluster.  If there is mass segregation, the
near-IR virial measurement probes only the core of the cluster, i.e.,
mass contained within the volume populated by these highest-mass
stars.  As such, the derived mass would be a lower limit and the IMF
of the entire cluster may follow the Kroupa form.  In this case, the
IMF measured in the core would appear ``top-heavy'' because of mass
segregation.  A nearby example of this effect is the young, massive
cluster R136 in the 30 Doradus nebula in the Large Magellanic Cloud
(LMC).  \citet{brandl96} found that the mass function in R136 steepens
with increasing distance from the cluster center, indicating strong
mass segregation.

At the adopted cluster age, stars more massive than $\sim 8$~M$_\odot$
have exploded as supernovae, and stars in the 6--8 M$_\odot$ range
have evolved off the main sequence \citep{schaller92}.  If we assume
that the core consists solely of stars that are (and remnants of
progenitors which were) larger than 2 M$_\odot$ as indicated by the
$L/M$ ratio, integration of the Kroupa IMF implies that the measured
core mass represents only one-third of the current cluster mass.  The
remaining stars (with masses smaller than 2 M$_\odot$) would be
distributed outside the core.  This distribution of stars would thus
require a total cluster mass of $\sim 2 \times 10^6$ M$_{\odot}$ to be
consistent with a Kroupa IMF for the population of the entire cluster.

Mass segregation is generally associated with the gradual
equipartition of energy via stellar encounters in old globular
clusters \citep{spitzer87}.  High mass stars sink to the center of a
cluster through dynamic interactions over the course of the relaxation
time, typically of order $10^8$ years for a globular cluster.  We do
not expect M82-F to exhibit mass segregation over its full extent at
its adopted age.  If we assume an average stellar mass of 0.87~M$_{\odot}$
for a full Kroupa IMF, the half-mass relaxation time \citep{meylan87}
for M82-F is $t_{rh} = 4 \times 10^8$ years --- roughly an order of
magnitude longer than the cluster's current age.

Recent studies, however, have found evidence of mass segregation in
significantly younger clusters.  \citet{lynne98} found that the
highest mass stars in the 0.8 Myr-old Orion Nebula Cluster are
preferentially located in the cluster center, and that stars down to
0.3 $M_{\odot}$ are less centrally concentrated than more massive
stars within the inner 1.0 pc.  The massive cluster R136 is mass
segregated at an age of only 3 Myr \citep{brandl96}.  Less massive LMC
clusters NGC 1805 \citep{degrijs02b}, NGC 1818, NGC 2004 and NCG 2100
\citep{gouliermis04} as well as NGC 330 \citep{sirianni02} in the
Small Magellanic Cloud all display mass segregation at ages of 10--50
Myr.  

Numerical simulations demonstrate that segregation of a cluster's most
massive stars occurs much more rapidly than the half-mass relaxation
time \citep{gerhard00}.  Indeed, if high mass stars are somehow
concentrated at the center of the cluster at the time of formation,
the relaxation timescale there will be shorter.  Dynamical mass
segregation will thus proceed more rapidly at the core, on the order
of a few crossing times \citep{degrijs02b}.  Based on the measured
velocity dispersion and half-light radius, the crossing time for M82-F
is $t_c = 1.4 \times 10^5$ years.  This is significantly less than the
age of the cluster, which implies that the core has had time to
undergo mass segregation.  \citet{degrijs02a} showed that the cores of
SSCs may undergo significant dynamical evolution in as little as 25
Myr.  The young age of M82-F relative to its half-mass relaxation time
may therefore imply rapid mass segregation for the massive stars or
support aggregation \citep{bonnell02} as a mechanism for high mass
star formation.

\section{Summary}

We present new, high resolution Keck/NIRSPEC echelle spectrometry
and new HST/ACS imaging of the super--star cluster M82-F.  The main
results of this study are:

\begin{enumerate}

\item Measuring the stellar velocity dispersion and half-light radius
of the cluster in the $H$-band, we find a virial mass of $6.6 \pm 0.9
\times 10^5$ M$_{\odot}$ within the half-light radius.  In the
$I$-band we find a mass of $7.0 \pm 1.2 \times 10^5$ M$_{\odot}$,
which supersedes the larger value quoted by SG01.

\item The cluster's light-to-mass ratio in the $I$- and $H$-bands are
inconsistent with a Kroupa IMF over the full range of stellar masses
from 0.1 to 100 M$_{\odot}$.  This may be due to a top-heavy IMF, with
a lower mass cutoff of about 2 M$_{\odot}$.

\item The cluster is successively larger at each shorter wavelength
imaged.  We interpret this result as due to mass segregation, even at
cluster's young age of about 50 Myr.

\item Mass segregation tends to overstate the $L/M$ ratio, as our
observations of $H$-band light are insensitive to the mass of stars at
radii larger than the volume occupied by red supergiant stars.  As
such, the measured mass represents a lower limit on the mass of the
entire cluster.

\item The IMF of M82-F differs from the IMF of nearby SSC MGG-9.  The
IMF in the nuclear starburst of M82 thus varies from cluster to
cluster.  This may be due to either IMF variation that is inconsistent
with a single, ``universal'' form or the degree of mass segregation
within individual clusters.

\end{enumerate}

%%%%%%%%%%%%%%%%%%%
% Acknowledgments %
%%%%%%%%%%%%%%%%%%%

\acknowledgments

We would like to thank the staff of the Keck Observatory, and
observing assistant Gary Puniwai specifically.  We also thank
L. J. Smith for providing us with optical spectroscopy of M82-F.  NM
thanks I. R. King for helpful discussions on elliptical clusters,
A. Cotton-Clay for mathematical assistance, and J. Terrell for
generous computing support.  The authors wish to recognize and
acknowledge the very significant cultural role and reverence that the
summit of Mauna Kea has always had within the indigenous Hawaiian
community.  We are most fortunate to have the opportunity to conduct
observations from this mountain.  This work has been supported by NSF
Grant AST--0205999.

%%%%%%%%%%%%%%
% References %
%%%%%%%%%%%%%%

\bibliographystyle{apj}
\bibliography{apj-jour,astro_refs}

%%%%%%%%%%%
% Figures %
%%%%%%%%%%%

\clearpage

%------------%
% IR Spectra %
%------------%

%----------%
% CCF plot %
%----------%

%-----------------%
% data, fit plots %
%-----------------%

%--------------%
% radius vs wl %
%--------------%

%-----------%
% L/M plots %
%-----------%

\end{document}